\documentclass[11pt]{article}
\usepackage{amssymb,amsmath,graphicx}
\usepackage{amsfonts,setspace,physics}
\usepackage{caption}
\usepackage[caption=false]{subfig}
\usepackage[numbers,sort,compress]{natbib}
\usepackage{epsfig,latexsym,graphicx,color}
\newtheorem{theorem}{Theorem}[section]

\begin{document}

\title{Optimal Guessing under Nonextensive Framework \\and associated Moment Bounds}

\author{
Abhik Ghosh \\
\small Interdisciplinary Statistical Research unit\\
\small Indian Statistical Institute, Kolkata, India  \\
\small{\it abhik.ghosh@isical.ac.in
}}
\maketitle

\begin{abstract}
We consider the problem of guessing the realization of a random variable 
but under more general Tsallis' non-extensive entropic framework rather than 
the classical Maxwell-Boltzman-Gibbs-Shannon framework. 
We consider both the conditional guessing problem in the presence of some related side information, 
and the unconditional one where no such side-information is available. 
For both types of the problem, the non-extensive moment bounds of the required number of guesses are derived;
here we use the $q$-normalized expectation in place of the usual (linear) expectation to define the non-extensive moments.
These moment bounds are seen to be a function of the logarithmic norm entropy measure, 
a recently developed two-parameter generalization of the Renyi entropy, 
and hence provide their information theoretic interpretation.
We have also considered the case of uncertain source distribution and derived the non-extensive moment bounds 
for the corresponding  mismatched guessing function. These mismatched bounds are interestingly seen to be linked with 
an important robust statistical divergence family known as the relative $(\alpha,\beta)$-entropies;
similar link is discussed between the optimum mismatched guessing  with the extremes of these relative entropy measures.
\end{abstract}


\section{Introduction}
\label{SEC:concept}

The problem of guessing the realization of a random variable is an well-known and important problem in  information theory,
motivated by the need of decoding a cryptic message in the output of a communication channel \citep{Arikan:1994,Massey:1994}.
Suppose $X$ be a random variable taking values in a finite set, say $\mathcal{X}$, 
and the probability mass function (pmf) of $X$ is $P_X(x)$.
We want to guess its realization by asking the question ``Is $X=x$?", varying $x\in\mathcal{X}$ sequentially,
until the answer is ``Yes". For any guessing strategy, 
let $G(x)$ denotes the number of guess required to reach the correct conclusion given $X=x$.
The optimal guessing strategy, obtained by minimizing $E(G(X))$, is to guess in the decreasing order of probabilities 
$\{P_X(x):x\in\mathcal{X}\}$; the minimum possible value of $E(G(X))$ is further related to 
the Shannon entropy of $X$ \cite{Shannon:1948} defined as 
\begin{eqnarray}
\mathcal{E}(X)=\mathcal{E}(P_X) = - \sum_{x\in\mathcal{X}} P_X(x)\ln P_X(x).
\label{EQ:Shannon_entropy}
\end{eqnarray}
If additionally a correlated random variable $Y$, taking values in a countable set $\mathcal{Y}$, 
is available, the objective becomes guessing $X$ for a given values of $Y=y$ by a guessing strategy $G(X|Y=y)$ 
and the best candidate turns out to be guessing in order of decreasing probabilities of $X$ given $Y=y$, say $P_{X|Y}(x|y)$;
see \cite{Massey:1994,Arikan:1994,Arikan:1996}.

Arikan \cite{Arikan:1996} have further extended the above guessing theory by considering the minimization 
of the moments of $G(X)$ and providing a tight bound in terms of the Renyi entropy measure \cite{Renyi:1961}.   
In particular, it has been shown in \cite{Arikan:1996} that, for any $\rho>0$,
\begin{eqnarray}
\mathcal{E}_{\frac{1}{(1+\rho)}}(P_X) - \ln(1+|\mathcal{X}|) \leq \frac{1}{\rho} \ln \left(\min_G E[G(X)^\rho]\right) 
\leq \mathcal{E}_{\frac{1}{(1+\rho)}}(P_X),
\label{EQ:moment_bound}
\end{eqnarray} 
where $\mathcal{E}_\alpha(P_X)$ denotes the Renyi entropy of order $\alpha$ of the distribution $P_X$ (or the underlying random variable  $X$) given by 
\begin{eqnarray}
\mathcal{E}_{\alpha}(P_X) := \frac{1}{1 - \alpha}\log \left[\sum_{x\in\mathcal{X}} P_X(x)^{\alpha}\right],
~~~ \alpha> 0.
\label{EQ:Renyi_entropy}
\end{eqnarray}
Note that $\mathcal{E}_{1}(P_X)$ is defined only in the limiting sense as $\alpha\rightarrow 1$
and coincides with the Shannon entropy $\mathcal{E}(P_X)$ given in (\ref{EQ:Shannon_entropy}).
For the conditional guessing given $Y$, moment bounds also similar to (\ref{EQ:moment_bound}) 
are studied in \cite{Arikan:1996}.
But all these results assume that the true source distribution $P_X$ or $P_{X|Y}$ is known in the respective cases.

The above guessing theory has been extended more recently by \cite{Sundaresan:2007,Kumar/Sundaresan:2015a} 
to the cases of uncertain sources, where the guesser only know that the source distribution 
is coming from a family $\mathcal{P}$ of pmfs over $\mathcal{X}$. 
Considering the conditional guessing problem with joint distribution of $(X,Y)$ denoted by $P$,
here one need to minimize the worst (supremum) value of the penalty or redundancy measure 
\begin{eqnarray}
R(P,G) =\frac{1}{\rho} \log E[G(X)^\rho] - \frac{1}{\rho} \log E[G_P(X)^\rho],
\label{EQ:penalty}
\end{eqnarray}
where $G_P$ denote the optimal guessing strategy when source has distribution $P$. 
The supremum and minimum of $R(P,G)$ should taken, respectively, over $P\in \mathcal{P}$ and all guessing strategy $G$.
The final optimum value $[\min_G\sup_P R(P,G)]$ in this case of uncertain source 
also satisfies a similar moment bound as in (\ref{EQ:moment_bound}) with Renyi entropy replaced by 
the corresponding relative entropy measure $\mathcal{RE}_\alpha(P, Q_G)$,
where $Q_G$ is a distribution on $\mathcal{X}$ such that $G_{Q_G}=G$;
see \cite{Sundaresan:2007} for details.

Most, if not all, works on guessing are developed with the ordinary (linear) expectation,
the basis of Shannon theory related to the classical Maxwell-Boltzmann-Gibbs (MBG) statistical physics.
However, more recently several complex systems are observed where the prediction of the MBG theory fails
leading to more general entropies and corresponding statistical frameworks. 
A popular extension is the Tsallis entropy \cite{Tsallis:1988} and associated  non-extensive statistics 
which are applied successfully to predict the behavior of many complex systems;
see, for example, \cite{Tsallis/etc:1998, Tsallis/etc:2005,Tsallis:2009, Majhi:2017,	Liu/Goree:2008, Devoe:2009}
and the references therein. These works lead to a whole new framework of non-extensive statistical physics 
which has also been applied to the information science generalizing the classical results of Shannon Coding theory; 
see \cite{Chapeau-Blondeau/etc:2011,Dukkipati/etc:2005,Suyari:2004,Plastino/Plastino:1999,Bialek/etc:2001,Martins/etc:2009, 
	Borland/etc:1998,Navarra/etc:2003,Yamano:2001,Yamano:2002} among many others. 

In this paper, we will extend the guessing theory and related moment inequalities under the Tsallis' 
non-extensive frameworks, considering suitably defined generalized expectation in place of the usual (linear) expectation. 
After some debates \cite{Tsallis:1994,Tsallis/etc:1998,Tsallis/etc:2005}, 
it is finally well accepted that the ``best" choice of constraints under the non-extensive framework
should be given in terms of the $q$-normalized expectation defined as 
\begin{eqnarray}
E_q[G(X)]=\frac{\sum_{x\in\mathcal{X}} G(x)P(x)^q}{\sum_{x\in\mathcal{X}} P(x)^q},
~~~ q\in\mathbb{R},
\label{EQ:q-NormExp}
\end{eqnarray}
for any function $G(\cdot)$ of $X$ (including the guessing function considered above).
Note that, these expectations can also be written as the linear expectation 
with respect to the $q$-escort distribution $P_q = P^q/W_q(P)$, with $W_q(P)=\sum_{x\in\mathcal{X}} P(x)^q$,
which has its own importance and applications in information theory 
(\cite{Pennini/etc:2007,Bercher:2009,Tanaka:2010,Abe:2003, Beck:2004,Bercher:2009}).
So, it is natural to study the guessing inequalities and the optimal guessing results of \cite{Arikan:1996,Sundaresan:2007} 
in terms of the non-extensive $q$-normalized expectation in (\ref{EQ:q-NormExp}), 
which is the main objective of the present paper.

Major contribution of this paper can be summarized as follows. 
\begin{itemize}
	\item We discuss the optimal guessing strategy, both for unconditional and conditional guessing problems, 
	obtained by minimizing the $q$-normalized moments of the number of guesses under Tsalli's non-extensive framework,
	and obtain the moments bound for the resulting optimum guessing function. 
	
	\item We obtain a lower bound of the non-extensive moments of the number of guesses required to correctly predict 
	a realization of a discrete random variable $X$ with and without additional side-information.
	The obtained bound is shown to be tight up to a multiplicative constant for the optimum strategy.
	
	\item We provide an information theoretic justification of a newly developed two-parameter 
	family of entropy measures, namely the logarithmic norm entropy (LNE), 
	which were developed as a generalization of Renyi entropy family.
	In this paper, we have proved a direct connection of the LNE measures 
	with the moment bound of the optimum guessing under non-extensive framework,
	indicating a new interpretation of these LNE measures. 
	As a by-product, we further extend the LNE measures to define the corresponding conditional entropy as well.
	
	\item We have also considered the cases where the source distribution is not exactly known and the guessing is to be done based 
	on a mismatched distribution. For both the unconditional and conditional problems, 
	we have derived the $q$-normalized moment bounds for the mismatched guessing functions under non-extensive framework. 
	The bound is again shown to be tight up to a multiplicative constant for the optimum strategy. 
	
	\item The moment bounds for non-extensive mismatched guessing are further shown to be linked with 
	the relative $(\alpha, \beta)$-entropy measure, also known as the logarithmic super divergences.
	These divergences were observed to be extremely useful in robust statistical inference \cite{Maji/etc:2014,Maji/etc:2016};
	we provide their information theoretic interpretation from mismatched guessing under non-extensivity.
	
	\item Finally we illustrate that the optimum guessing strategy with mismatched source under non-extensivity 
	can be obtained by minimizing the maximum of the relative $(\alpha, \beta)$-entropies 
	between the mismatched source and all plausible true source distributions. 
	Non-extensive moment bounds for the resulting optimum guessing function is also derived.
\end{itemize}

\section{Optimal Guessing via Non-Extensive Moment Criterion}
\label{SEC:Moments}

\subsection{Bounds on the Non-Extensive Moments of the Number of Guesses}

Consider the problem of guessing the realization of the random variable $X$ with finite support $\mathcal{X}$
along with the notation of Section \ref{SEC:concept}. We start with proving an important inequality 
on the non-extensive $q$-normalized moments of $G(X)$.

\begin{theorem}
\label{THM:Moment_bound_X}
For any arbitrary guessing function $G(X)$, any $\rho > 0$ and any $q\in \mathbb{R}$, we have	
\begin{eqnarray}
E_q[G(X)^\rho] \geq 
(1+\ln|\mathcal{X}|)^{-\rho} \frac{\left[\sum_{x\in\mathcal{X}}P_X(x)^{\frac{q}{1+\rho}}\right]^{1+\rho}}{
	\sum_{x\in\mathcal{X}}P_X(x)^q}.
\label{EQ:moment_boundNE_X}
\end{eqnarray}
\end{theorem}
\noindent\textbf{Proof:}\\
For simplicity let us drop the subscript in $P_X(x)$ and use the notation $W_q(P) = \sum\limits_{x\in\mathcal{X}}P(x)^q$. 
Now, taking an arbitrary distribution $Q$ on $\mathcal{X}$, we have
\begin{eqnarray}
E_q[G(X)^\rho] &=& \frac{\sum\limits_{x\in\mathcal{X}}G(x)^\rho P(x)^q}{W_q(P)}
=\sum_{x\in\mathcal{X}} Q(x) \exp\left[- \ln \frac{Q(x)W_q(P)}{G(X)^\rho P(x)^q}\right]
\nonumber\\
&\geq& \exp\left[-\mathcal{RE}(Q,P) + \rho \sum_{x\in\mathcal{X}}Q(x)\ln G(x) + (q-1)\sum_{x\in\mathcal{X}}Q(x)\ln P(x) - \ln W_q(P)\right],
~~~~
\label{EQ:2.1}
\end{eqnarray}
by the application of Jensen's inequality, where $\mathcal{RE}(Q,P) $ denotes 
the Kullback-Leibler relative entropy measure \cite{Kullback/Leibler:1951} defined as
\begin{eqnarray}
\mathcal{RE}(Q,P) =\sum_{x\in\mathcal{X}} Q(x)\ln\frac{Q(x)}{P(x)}.
\label{EQ:KLD}
\end{eqnarray}
Now, in terms of $\mathcal{E}(P)$ from (\ref{EQ:Shannon_entropy}), we get
\begin{eqnarray}
\sum_{x\in\mathcal{X}} Q(x)G(x) = \mathcal{E}(Q) - \sum_{x\in\mathcal{X}} Q(x) \ln \frac{1}{Q(x)G(x)}
\geq \mathcal{E}(Q) - \ln\sum_{x\in\mathcal{X}} \frac{1}{G(x)},
\nonumber
\end{eqnarray}
by another application of Jensen's inequality. But, we know
$$
\sum_{x\in\mathcal{X}}\frac{1}{G(x)} = \sum_{i=1}^{|\mathcal{X}|} \frac{1}{i} \leq 1+ \ln|\mathcal{X}|.
$$
Hence, combining above equations, we get 
\begin{eqnarray}
\sum_{x\in\mathcal{X}} Q(x)G(x) \geq \mathcal{E}(Q) - \ln(1+ \ln|\mathcal{X}|).
\label{EQ:2.2}
\end{eqnarray}
Further, simple algebra yields
\begin{eqnarray}
\sum_{x\in\mathcal{X}} Q(x)\ln P(x) = - \mathcal{E}(Q) - \mathcal{RE}(Q,P).
\label{EQ:2.3}
\end{eqnarray}
Now, substituting (\ref{EQ:2.2}) and (\ref{EQ:2.3}) in (\ref{EQ:2.1}), we get
\begin{eqnarray}
E_q[G(X)^\rho] 
&\geq& (1+\ln|\mathcal{X}|)^{-\rho} \exp\left[-q\mathcal{RE}(Q,P) + (\rho-q+1) \mathcal{E}(Q) - \ln W_q(P)\right].
\label{EQ:2.4}
\end{eqnarray}
Finally, by the standard Lagrange multiplier arguments, one can show that, given $P$, the quantity
$\left[(\rho-q+1) \mathcal{E}(Q) - q\mathcal{RE}(Q,P)\right]$ is maximized over $Q$ 
at the distribution 
\begin{eqnarray}
Q^\ast(x) = \frac{P(x)^\frac{q}{1+\rho}}{\sum_{x'\in\mathcal{X}}P(x')^\frac{q}{1+\rho}}, ~~~~~x\in \mathcal{X}.
\end{eqnarray}
Hence, a tight bound of $E_q[G(X)^\rho]$ can be obtained by substituting the above choice of $Q=Q^\ast$ in (\ref{EQ:2.4}),
which leads to the desired result (\ref{EQ:moment_boundNE_X}). 
\hfill{$\square$}

\bigskip
Next we consider the conditional guessing problem with notation of Section \ref{SEC:concept}
and derive the lower bound on the non-extensive moment of $G(X|Y)$ which is presented in the following theorem. 
Here we denote the joint pmf of $(X,Y)$ by $P_{X,Y}(x,y)$ and the marginal pmf of $Y$ by $P_Y(y)$;
note that $P_{X,Y}(x,y) = P_{X|Y}(x|y)P_Y(y)$.

\begin{theorem}
	\label{THM:Moment_bound_XY}
	For any arbitrary conditional guessing function $G(X|Y)$, any $\rho > 0$ and any $q\in \mathbb{R}$, we have	
	\begin{eqnarray}
E_q[G(X|Y)^\rho] \geq (1+\ln|\mathcal{X}|)^{-\rho}
\frac{\sum_{y\in\mathcal{Y}}\left[\sum_{x\in\mathcal{X}}P_{X,Y}(x,y)^{\frac{q}{1+\rho}}\right]^{1+\rho}}{
		\sum_{y\in\mathcal{Y}}\sum_{x\in\mathcal{X}}P_{X,Y}(x,y)^q}.
	\label{EQ:moment_boundNE_XY}
	\end{eqnarray}
\end{theorem}
\noindent\textbf{Proof:}\\
We break the joint probability into conditional and marginal probabilities, 
and apply Theorem \ref{THM:Moment_bound_X} to the conditional expectation, to get
\begin{eqnarray}
E_q[G(X|Y)^\rho] 
&=& \frac{\sum_{x\in\mathcal{X}}\sum_{y\in\mathcal{Y}}P_{X,Y}(x,y)^qG(x|Y=y)}{
	\sum_{x\in\mathcal{X}}\sum_{y\in\mathcal{Y}}P_{X,Y}(x,y)^q}
\nonumber\\
&=& \frac{\sum_{x\in\mathcal{X}}\sum_{y\in\mathcal{Y}}P_{X|Y}(x|y)^qP_Y(y)^qG(x|Y=y)}{
	\sum_{x\in\mathcal{X}}\sum_{y\in\mathcal{Y}}P_{X|Y}(x|y)^qP_Y(y)^q}
\nonumber\\
&=& \frac{\sum_{y\in\mathcal{Y}}P_Y(y)^q\left\{E_q[G(X|Y=y)|Y=y] \left(\sum_{x\in\mathcal{X}}P_{X|Y}(x|y)^q\right)\right\}}{
	\sum_{x\in\mathcal{X}}\sum_{y\in\mathcal{Y}}P_{X|Y}(x|y)^qP_Y(y)^q}
\nonumber\\
&\geq& (1+\ln|\mathcal{X}|)^{-\rho}
\frac{\sum_{y\in\mathcal{Y}}P_Y(y)^q\left[\sum_{x\in\mathcal{X}}P_{X|Y}(x|y)^{\frac{q}{1+\rho}}\right]^{1+\rho}}{
	\sum_{y\in\mathcal{Y}}\sum_{x\in\mathcal{X}}P_{X|Y}(x|y)^qP_Y(y)^q}.
\nonumber\\
&=& (1+\ln|\mathcal{X}|)^{-\rho}
\frac{\sum_{y\in\mathcal{Y}}\left[\sum_{x\in\mathcal{X}}P_{X,Y}(x,y)^{\frac{q}{1+\rho}}\right]^{1+\rho}}{
	\sum_{y\in\mathcal{Y}}\sum_{x\in\mathcal{X}}P_{X,Y}(x,y)^q}.\nonumber
\end{eqnarray}
This proves the theorem.
\hfill{$\square$}

\bigskip
It is interesting to note another interpretation of these lower bounds obtained in above two theorems
through the escort distribution. Let us denote by $P_q(x,y)$, $P_q(x|y)$ and $P_q(x)$  the escort distributions 
corresponding to the joint, conditional and marginal pmfs $P_{X,Y}(x,y)$, $P_{X|Y}(x|y)$ and $P_X(x)$, respectively.
Then we can rewrite the main results of the previous two theorems as 
\begin{eqnarray}
E_q[G(X)^\rho] 
&\geq& (1+\ln|\mathcal{X}|)^{-\rho}
{\left[\sum_{x\in\mathcal{X}}P_{q}(x)^{\frac{1}{1+\rho}}\right]^{1+\rho}},
\label{EQ:esc_bnd1}\\
E_q[G(X|Y)^\rho] &\geq& (1+\ln|\mathcal{X}|)^{-\rho}
{\sum_{y\in\mathcal{Y}} P_q(\cdot, y)\left[\sum_{x\in\mathcal{X}}P_{q}(x|y)^{\frac{1}{1+\rho}}\right]^{1+\rho}},
\label{EQ:esc_bnd2}
\end{eqnarray}
where $P_q(\cdot, y)=\sum_{x\in\mathcal{X}}P_{q}(x,y)$ is the marginal escort distribution of $Y$.
Note that, at $q=1$, all escort distributions coincide with the respective origin distributions
and $P_1(\cdot, y) = P_Y(y)$; hence our results coincide with those of \cite{Arikan:1996} at $q=1$.
It links our results with classical ones through the concept of escort distribution under non-extensive framework.

Further, the above lower bounds in (\ref{EQ:esc_bnd1}) and (\ref{EQ:esc_bnd2})  
are valid for any guessing function, not necessarily the optimal one.
In the following subsection, we will define the optimal guessing strategy 
and develop a complementary upper bound of the non-extensive moments of the optimum number of guesses.

\subsection{Optimal Guessing under Non-Extensivity}

Let us first consider the conditional guessing problem. We call a guessing strategy $G(X|Y)$ to be 
\textit{optimal under $q$-non-extensivity} if it minimizes the non-extensive moments $E_q[G(X|Y)^\rho]$ 
simultaneously for all $\rho > 0$. Note that, in terms of the escort distributions, we can write
\begin{eqnarray}
E_q[G(X|Y)^\rho] = {\sum_{y\in\mathcal{Y}} P_q(\cdot, y)\sum_{x\in\mathcal{X}}P_{q}(x|y)G(x|y)^\rho},
\label{EQ:q-exp1}
\end{eqnarray}
which is minimized by the guessing function $G^\ast(X|Y)$ satisfying 
\begin{eqnarray}
G^\ast(x|y)<G^\ast(x,|y) ~~~\Rightarrow ~~ P_q(x|y) \geq P_q(x'|y), ~~~\mbox{ for all }~x, x'\in\mathcal{X}, ~y\in\mathcal{Y}.
\label{EQ:opt_guess}
\end{eqnarray}
Therefore, the optimal guessing rule $G^\ast(X|Y)$ is to guess the values of $X$, given $Y=y$,
in decreasing order of the $q$-escort distribution $P_q(x|y)$ 
of the conditional (posterior) pmf$P_{X|Y}(x|y)$. 
This optimal guessing rule is unique if and only if $P_q(x|y)$ or equivalently $P_{X|Y}(x|y)$ 
is distinct over $x\in \mathcal{X}$ for any given $Y=y$; this is exactly the same uniqueness condition 
as in the case of classical optimal guessing strategy of \cite{Arikan:1996}.

Next note that, we already have the lower bound of the optimal guessing function $G^\ast(X|Y)$ 
from Theorem \ref{THM:Moment_bound_XY}.
The following theorem presents its upper bound which is tight within a multiplicative factor of the lower bound. 

\begin{theorem}
	\label{THM:Moment_bound_opt}
	For the optimal guessing function $G^\ast(X|Y)$ under $q$-non-extensivity with any $q\in \mathbb{R}$
	and for any $\rho > 0$, we have	
	\begin{eqnarray}
	E_q[G^\ast(X|Y)^\rho] &\leq& 
	{\sum_{y\in\mathcal{Y}} P_q(\cdot, y)\left[\sum_{x\in\mathcal{X}}P_{q}(x|y)^{\frac{1}{1+\rho}}\right]^{1+\rho}}
	\nonumber\\
	&=&\frac{\sum_{y\in\mathcal{Y}}\left[\sum_{x\in\mathcal{X}}P_{X,Y}(x,y)^{\frac{q}{1+\rho}}\right]^{1+\rho}}{
		\sum_{y\in\mathcal{Y}}\sum_{x\in\mathcal{X}}P_{X,Y}(x,y)^q}.
	\label{EQ:moment_boundNE_opt}
	\end{eqnarray}
\end{theorem}
\noindent\textbf{Proof:}\\
We know that the optimal rule $G^\ast(x|y)$ satisfies (\ref{EQ:opt_guess})
and hence we have 
\begin{eqnarray}
G^\ast(x|y)&=& \sum_{x':G^\ast(x'|y)\leq G^\ast(x|y)}1 
\leq \sum_{x':G^\ast(x'|y)\leq G^\ast(x|y)}\left(\frac{P_q(x'|y)}{P_q(x|y)}\right)^{\frac{1}{1+\rho}}
\nonumber\\
&\leq&  \sum_{x'\in\mathcal{X}}\left(\frac{P_q(x'|y)}{P_q(x|y)}\right)^{\frac{1}{1+\rho}}.
\end{eqnarray}
Therefore, we get
\begin{eqnarray}
E_q[G^\ast(X|Y)^\rho] 
&=& {\sum_{y\in\mathcal{Y}} P_q(\cdot, y)\sum_{x\in\mathcal{X}}P_{q}(x|y)G^\ast(x|y)^\rho},
\nonumber\\
&\leq & {\sum_{y\in\mathcal{Y}} P_q(\cdot, y)\sum_{x\in\mathcal{X}}P_{q}(x|y)\left[\sum_{x'\in\mathcal{X}}
	\left(\frac{P_q(x'|y)}{P_q(x|y)}\right)^{\frac{1}{1+\rho}}\right]^\rho},
\nonumber\\
&=& {\sum_{y\in\mathcal{Y}} P_q(\cdot, y)\left[\sum_{x\in\mathcal{X}}P_{q}(x|y)^{\frac{1}{1+\rho}}\right]^{1+\rho}}.
\nonumber
\end{eqnarray}
The second part follows by straightforward algebra using the definitions of escort distributions. 
\hfill{$\square$}

\bigskip
Let us now identify and characterize the bound on the $q$-non-extensive moments of 
the optimal guessing function $G^\ast(X|Y)$ which is given by the right hand side of (\ref{EQ:moment_boundNE_opt});
let us denote this quantity as $L_{q,\rho}(X|Y)$. 
This bound is tight up to a multiplicative factor of $(1+\ln M)^\rho$, i.e.,
\begin{eqnarray}
(1+\ln|\mathcal{X}|)^{-\rho} L_{q,\rho}(X|Y)
\leq E_q[G^\ast(X|Y)^\rho] 
\leq L_{q,\rho}(X|Y).
\label{EQ:moment_inq_Cond}
\end{eqnarray}

In a similar manner, one can also deduce a similar tight bound for the unconditional guessing problem. 
The \textit{optimal guessing rule} $G^\ast(X)$ to guess the values of $X$ under \textit{$q$-non-extensivity},
defined by the simultaneous minimizer of  the non-extensive moments $E_q[G(X|Y)^\rho]$ for all $\rho > 0$,
is given by the decreasing order of the $q$-escort distribution $P_q(x)$ of $X$ 
and satisfies the moment inequality
\begin{eqnarray}
(1+\ln|\mathcal{X}|)^{-\rho} L_{q,\rho}(X)
\leq E_q[G^\ast(X)^\rho] 
\leq L_{q,\rho}(X),
\label{EQ:moment_inq}
\end{eqnarray}
where the bound $L_{q,\rho}(X)$ is the one in Theorem \ref{THM:Moment_bound_X}, i.e.,
\begin{eqnarray}
L_{q,\rho}(X)=\frac{\left[\sum_{x\in\mathcal{X}}P_X(x)^{\frac{q}{1+\rho}}\right]^{1+\rho}}{
	\sum_{x\in\mathcal{X}}P_X(x)^q}={\left[\sum_{x\in\mathcal{X}}P_{q}(x)^{\frac{1}{1+\rho}}\right]^{1+\rho}}.
\label{EQ:bound}
\end{eqnarray}

\bigskip\noindent
\textit{Relation of the bounds with a generalization of Renyi Entropy:}\\
It is interestingly to note that the above bounds $L_{q,\rho}(X)$ and $L_{q,\rho}(X|Y)$ are directly linked with a recent generalized entropy measure,
namely the logarithmic norm entropy (LNE) of \cite{Ghosh/Basu:2019}. 
The LNE of the distribution $P_X$ of X is defined in terms of two parameters $\alpha, \beta$ as 
\begin{eqnarray}
\mathcal{E}_{(\alpha, \beta)}(X)=\mathcal{E}_{(\alpha, \beta)}(P_X) = \frac{\alpha\beta}{(\beta-\alpha)}
\ln\frac{\left(\sum_{x\in\mathcal{X}}P_X(x)^\alpha\right)^{1/\alpha}}{\left(\sum_{x\in\mathcal{X}}P_X(x)^\beta\right)^{1/\beta}},
~~~~~~\alpha~>0, \beta\in \mathbb{R}\setminus\{\alpha\}.
\label{EQ:LNE}
\end{eqnarray}
It coincides with the classical Renyi entropy measure (\ref{EQ:Renyi_entropy}) if either of the two parameters equals one
and hence provides a two parameter generalization of Renyi entropy. The one parameter subclass at $\alpha=\beta$ 
is defined in the limiting sense and includes the Shannon entropy (\ref{EQ:Shannon_entropy})
at $\alpha=\beta=1$; see \cite{Ghosh/Basu:2018,Ghosh/Basu:2019} for more details.

By a simple algebra, one can easily see that the bound $L_q(X)$ in the unconditional problem is indeed given by 
\begin{eqnarray}
\ln L_{q,\rho}(X)= \rho \mathcal{E}_{(\frac{q}{1+\rho},q)}(X).
\end{eqnarray}
Then, the moment inequality in (\ref{EQ:moment_inq}) can be rewritten as 
\begin{eqnarray}
 \mathcal{E}_{(\frac{q}{1+\rho},q)}(X) - \ln(1+\ln|\mathcal{X}|)
\leq \frac{1}{\rho} \ln E_q[G^\ast(X)^\rho] 
\leq \mathcal{E}_{(\frac{q}{1+\rho},q)}(X).
\label{EQ:moment_inqE}
\end{eqnarray}
This provides a new interesting interpretation of the newly proposed LNE measure
through the non-extensive information theory, as well as the corresponding optimal guessing.

A similar interpretation of the conditional moment bound $L_{q,\rho}(X|Y)$ can also be obtained
if we extend the definition of the LNE measure to define the Conditional logarithmic norm entropy (CLNE) measure as
\begin{eqnarray}
\mathcal{E}_{(\alpha, \beta)}(X|Y)=\mathcal{E}_{(\alpha, \beta)}(P_X|P_Y) = \frac{\alpha}{(\beta-\alpha)}
\ln\frac{\sum_{y\in\mathcal{Y}}\left(\sum_{x\in\mathcal{X}}P_{X,Y}(x,y)^\alpha\right)^{\frac{\beta}{\alpha}}}{
	\sum_{y\in\mathcal{Y}}\left(\sum_{x\in\mathcal{X}}P_{X,Y}(x,y)^\beta\right)},
~~~\alpha~>0, \beta\in \mathbb{R}\setminus\{\alpha\}.
\label{EQ:CLNE}
\end{eqnarray}
Then, we can derive that $\ln L_{q,\rho}(X|Y) = \rho \mathcal{E}_{(\frac{q}{1+\rho},q)}(X)$, 
and hence the moment inequality (\ref{EQ:moment_inq_Cond}) can be rewritten as
\begin{eqnarray}
\mathcal{E}_{(\frac{q}{1+\rho},q)}(X|Y) - \ln(1+\ln|\mathcal{X}|)
\leq \frac{1}{\rho} \ln E_q[G^\ast(X|Y)^\rho] 
\leq \mathcal{E}_{(\frac{q}{1+\rho},q)}(X|Y),
\label{EQ:moment_inqE_Cond}
\end{eqnarray}
i.e.,
\begin{eqnarray}
\mathcal{E}_{(\frac{q}{1+\rho},q)}(X|Y) - \ln(1+\ln|\mathcal{X}|)
\leq \frac{1}{\rho} \ln \left(\min\limits_{G} E_q[G(X|Y)^\rho] \right)
\leq \mathcal{E}_{(\frac{q}{1+\rho},q)}(X|Y).
\nonumber
\end{eqnarray}
This final equation generalizes Arikan's \cite{Arikan:1996} guessing theorem for the non-extensive expectation.
Further, along with providing the bound for optimal guessing, 
we additionally obtain a new two-parameter family of conditional entropy measure in (\ref{EQ:CLNE})
which coincides with the Renyi conditional entropy if either $\alpha$ or $\beta$ equals one.
We can further extend this LNE family at $\alpha=\beta$ through continuous limit
which yields
\begin{eqnarray}
\mathcal{E}_{(\alpha, \alpha)}(X|Y)&=&\lim\limits_{\beta\rightarrow\alpha}\mathcal{E}_{(\alpha, \beta)}(X|Y) 
\nonumber\\
&=& -\alpha\frac{\sum_{y\in\mathcal{Y}}\left(\sum_{x\in\mathcal{X}}P_{X,Y}(x,y)^\alpha\ln P_{X,Y}(x,y)\right)}{
	\sum_{y\in\mathcal{Y}}\left(\sum_{x\in\mathcal{X}}P_{X,Y}(x,y)^\alpha\right)}
\nonumber\\
&& +\frac{\sum_{y\in\mathcal{Y}}\left(\sum_{x\in\mathcal{X}}P_{X,Y}(x,y)^\alpha\right)\ln\left(\sum_{x\in\mathcal{X}}P_{X,Y}(x,y)^\alpha\right)}{
	\sum_{y\in\mathcal{Y}}\left(\sum_{x\in\mathcal{X}}P_{X,Y}(x,y)^\alpha\right)}.
\label{EQ:CLNE0}
\end{eqnarray}
We hope to develop more interesting properties of these new CLNE measures in future works. 
An immediate property in the context of optimal guessing is obtained by taking limit as $\rho \rightarrow 0^{+}$ in (\ref{EQ:moment_inqE_Cond})
which gives
\begin{eqnarray}
\mathcal{E}_{(q, q)}(X|Y) = E_q[\ln G^\ast(X|Y)]  = \min_G E_q[\ln G(X|Y)]  .
\label{EQ:moment_eqE0_Cond}
\end{eqnarray}

\section{The Cases of Uncertain Source Distribution}
\label{SEC:Moments_uncertain}

We have studied the optimal guessing strategy and its non-extensive moments 
in the previous section, where we have assumed that the true joint distribution $P_{X,Y}(x,y)$ is known.
Let us now assume the case of uncertain source where the true distribution $P_{X,Y}(x,y)$ is not known
and it is only known that $P_{X,Y}$ comes from a family of probability distribution $\mathcal{P}$ 
over $\mathcal{X}\times\mathcal{Y}$. As noted in the introduction, in such a case
the optimal guessing strategy needs to be obtained by minimizing the worst (supremum) value of 
the penalty or redundancy measure $R(P,G)$ defined in (\ref{EQ:penalty});
however we will use the $q$-normalized expectations under the non-extensivity framework of the present paper.

Let us first focus on the conditional guessing problem, for which the $q$-non-extensive optimal guessing strategy $G^\ast(X|Y)$ 
is studied in the previous section; it guesses the values of $X$ given $Y=y$ in decreasing order of $P_q(x|y)$
obtained from $P_{X,Y}$. From now on, let us drop the subscript in $P_{X,Y}$ and 
denote the optimal strategy $G^\ast$ obtained from $P=P_{X,Y}$ by $G^\ast_P(X|Y)$.
However, due to the lack of knowledge, we can only guess based on another (joint) pmf $Q(x,y)$;
let us denote the corresponding guessing strategy by $G_Q^\ast(X|Y)$  which guesses the values of $X$ given $Y=y$ 
in decreasing order of $Q_q(x|y)$, the $q$-escort distribution of 
the conditional pmf $Q(x|y)=Q(x,y)/\int Q(x,y)dx$.
We start with deriving bounds for the non-extensive $q$-normalized expectation of the guessing function $G_Q^\ast(X|Y)$
under the true (but unknown) source distribution $P(x,y)$ in the following two theorems.

\begin{theorem}
\label{THM:Mismatch1}
Under the non-extensive conditional guessing problem with uncertain source, 
for any $\rho> 0$ and any $q\in \mathbb{R}$, we have
\begin{eqnarray}
E_q\left[G_Q^\ast(X|Y)^\rho\right] &\leq& {\sum_{y\in\mathcal{Y}} P_q(\cdot, y)\sum_{x\in\mathcal{X}}P_{q}(x|y)
\left[\sum_{x'\in\mathcal{X}}\left(\frac{Q_q(x'|y)}{Q_q(x|y)}\right)^{\frac{1}{1+\rho}}\right]^{\rho}},
\label{EQ:mismatch_bound1}
\end{eqnarray}
where the expectation is taken with respect to the joint distribution $P(x,y)$.
\end{theorem}
\noindent\textbf{Proof:}\\
The proof follows from the definition of $G_Q^\ast$ and (\ref{EQ:q-exp1})
by observing that 
\begin{eqnarray}
G_Q^\ast(x|y)&\leq& \sum_{x'\in\mathcal{X}}I\left[Q_q(x'|y)\geq Q_q(x|y)\right] 
\leq \sum_{x'\in\mathcal{X}}\left(\frac{Q_q(x'|y)}{Q_q(x|y)}\right)^{\frac{1}{1+\rho}}
\nonumber
\end{eqnarray}
\hfill{$\square$}

\begin{theorem}
\label{THM:Mismatch2}
Under the non-extensive conditional guessing problem, 
let $G(X|Y)$ denote any arbitrary guessing strategy and let $\rho>0$, $q\in \mathbb{R}\setminus\{0\} $. 
Then, there is a pmf $ Q^{(G)}$, depending on $G$, with support $\mathcal{X}\times\mathcal{Y}$ 
which satisfies
\begin{eqnarray}
E_q\left[G(X|Y)^\rho\right] &\geq& (1+\ln|\mathcal{X}|)^{-\rho} 
{\sum_{y\in\mathcal{Y}} P_q(\cdot, y)\sum_{x\in\mathcal{X}}P_{q}(x|y)
\left[\sum_{x'\in\mathcal{X}}\left(\frac{Q_{q}^{(G)}(x'|y)}{Q_q^{(G)}(x|y)}\right)^{\frac{1}{1+\rho}}\right]^{\rho}},
\label{EQ:mismatch_bound2}
\end{eqnarray}
where the expectation is taken with respect to the joint distribution $P(x,y)$,
and $Q_q^{(G)}(x|y)$ denotes the $q$-escort distribution of the conditional pmf $Q^{(G)}(x|y)=Q^{(G)}(x,y)/\int Q^{(G)}(x,y)dx$.
\end{theorem}
\noindent\textbf{Proof:}\\
Let us define, for $\rho>0$, $q\in\mathbb{R}-\{0\}$ and for each $y\in\mathcal{Y}$, 
$$
s_{\rho, q} := \sum_{x\in\mathcal{X}}\left(\frac{1}{G(x,y)}\right)^{\frac{1+\rho}{q}} = \sum_{i=1}^{|\mathcal{X}|}\frac{1}{i^{\frac{1+\rho}{q}}}.
$$ 
Note that, clearly $s_{\rho, q}$ is independent of $y\in\mathcal{Y}$ and is finite for all $\rho>0$ and $q\in\mathbb{R}-\{0\}$.
Given the guessing strategy $G(x,y)$, define the joint pmf $Q^{(G)}$ on $\mathcal{X}\times\mathcal{Y}$ as
$$
Q^{(G)}(x,y) = \frac{1}{|\mathcal{Y}|s_{\rho, q}G(x,y)^{\frac{1+\rho}{q}}},
~~~~~~ \mbox{ for all }~~(x,y)\in \mathcal{X}\times\mathcal{Y}.
$$
It is easy to verify that $Q^{(G)}$ is a joint pmf with support $\mathcal{X}\times\mathcal{Y}$ and
$$
Q_q^{(G)}(x|y) = \frac{1}{s_{\rho, 1}G(x,y)^{{1+\rho}}},
~~~~~~ \mbox{ for all }~~(x,y)\in \mathcal{X}\times\mathcal{Y}.
$$
Now, using the above formula $Q_q^{(G)}(x|y)$, we get
\begin{eqnarray}
&&{\sum_{y\in\mathcal{Y}} P_q(\cdot, y)\sum_{x\in\mathcal{X}}P_{q}(x|y)
	\left[\sum_{x'\in\mathcal{X}}\left(\frac{Q_{q}^{(G)}(x'|y)}{Q_q^{(G)}(x|y)}\right)^{\frac{1}{1+\rho}}\right]^{\rho}}
\nonumber\\
&&~~~~= {\sum_{y\in\mathcal{Y}} P_q(\cdot, y)\sum_{x\in\mathcal{X}}P_{q}(x|y)
	\left[\sum_{x'\in\mathcal{X}}\frac{G(x|y)}{G(x'|y)}\right]^{\rho}}
\nonumber\\
&&~~~~= \sum_{y\in\mathcal{Y}} P_q(\cdot, y)\sum_{x\in\mathcal{X}}P_{q}(x|y)G(x|y)^\rho s_{0,1}^\rho
= s_{0,1}^\rho E_q\left[G(X|Y)^\rho\right].
\end{eqnarray}
Then the theorem follows by noting that $s_{0,1}=\sum\limits_{i=1}^{|\mathcal{X}|}\frac{1}{i} \leq 1+ \ln|\mathcal{X}|$.
\hfill{$\square$}

\bigskip
Let us denote the right hand side of (\ref{EQ:mismatch_bound1}) by $L_{q,\rho}^\ast(P,Q)$.
Then, combining the results from Theorems \ref{THM:Mismatch1} and \ref{THM:Mismatch2}, 
we have the following non-extensive moment bound for mismatched guessing strategy
\begin{eqnarray}
(1+\ln|\mathcal{X}|)^{-\rho} L_{q,\rho}^\ast(P, Q^{(G_Q^\ast)})
\leq E_q[G_Q^\ast(X|Y)^\rho] 
\leq L_{q,\rho}^\ast(P,Q),
\label{EQ:moment_inq_Cond_mismatch}
\end{eqnarray}
where the expectation is taken with respect to the joint distribution $P(x,y)$.
Note that (\ref{EQ:moment_inq_Cond_mismatch}) complements (\ref{EQ:moment_inq_Cond})
for the cases of uncertain source; they coincide when the source is known, i.e., when $Q=P$.

To get physical interpretation of the above bounds, let us define 
$$
\mathcal{RE}_{(\alpha, \beta)}(P,Q) = \frac{\alpha}{\beta(\beta-\alpha)}\ln L_{\beta, \frac{(\beta-\alpha)}{\alpha}}^\ast(P,Q) 
- \frac{1}{\beta}\mathcal{E}_{( \alpha, \beta)}(P_X|P_Y), 
~~~~~~\alpha>0, \beta\in\mathbb{R}\setminus\{\alpha\}
$$
where the second term is as defined in (\ref{EQ:CLNE}) from $P$. 
After some algebra, one can simplify this measure to have the form
\begin{eqnarray}
\mathcal{RE}_{(\alpha, \beta)}(P,Q)
= \frac{\alpha}{\beta(\beta-\alpha)}\ln\frac{\sum\limits_{y\in\mathcal{Y}}\left\{\sum\limits_{x\in\mathcal{X}}P(x,y)^\beta 
	Q(x,y)^{\alpha-\beta}\right\}\left\{\sum\limits_{x\in\mathcal{X}}Q(x,y)^{\beta}\right\}^{\frac{\beta}{\alpha}-1}}{
\sum\limits_{y\in\mathcal{Y}}\left\{\sum\limits_{x\in\mathcal{X}}P(x,y)^\beta\right\}^{\frac{\beta}{\alpha}}}.
\end{eqnarray}
For the case when $|\mathcal{Y}|=1$, i.e., the case of no additional information to condition upon, 
our joint pmfs $P(x,y)$ and $Q(x,y)$ may be though of as the pmfs of $X$ only over $\mathcal{X}$, say $P_X(x)$ and $Q_X(x)$. 
In this case the above measure simplifies to
\begin{eqnarray}
&&\mathcal{RE}_{(\alpha, \beta)}(P_X,Q_X)
\nonumber\\
&=& \frac{1}{(\alpha-\beta)}\ln\left\{\sum_{x\in\mathcal{X}}P_X(x)^\beta\right\}
- \frac{\alpha}{\beta(\alpha-\beta)}\ln\left\{\sum_{x\in\mathcal{X}}P_X(x)^\beta Q_X(x)^{\alpha-\beta}\right\}
+\frac{1}{\beta}\ln\left\{\sum_{x\in\mathcal{X}}Q_X(x)^{\beta}\right\},~~~~~~~
\label{EQ:RE_entropy}
\end{eqnarray}
which is exactly the relative $(\alpha,\beta)$-entropy studied in \cite{Ghosh/Basu:2018}.
This provides a two-parameter generalization of the relative $\alpha$-entropy of \cite{Kumar/Sundaresan:2015a} 
or equivalently of the Renyi divergence family.
For all $\alpha>0$ and $\beta\in\mathbb{R}$, it has been shown that the relative $(\alpha,\beta)$-entropy
is indeed a proper statistical divergence and hence $\mathcal{RE}_{(\alpha, \beta)}(P,Q)\geq 0$ 
with equality if and only if $P=Q$ \cite{Ghosh/Basu:2018}.
This particular two parameter divergence family has further importance in robust statistical inference,
where it was referred to as the logarithmic super divergence family \citep{Maji/etc:2014,Maji/etc:2016}.

Recalling from (\ref{EQ:penalty}), let us now define our target redundancy measure 
for the conditional guessing problem under non-extensivity 
by using $q$-normalized expectation 
\begin{eqnarray}
R_q(P,G) =\frac{1}{\rho} \ln E_q[G(X|Y)^\rho] - \frac{1}{\rho} \ln E_q[G_P(X|Y)^\rho],
\label{EQ:penalty_NE}
\end{eqnarray}
where the expectation is taken with respect to the joint distribution $P(x,y)$.
We can obtain the bound on its value from (\ref{EQ:moment_inq_Cond_mismatch})
which is presented in the following theorem. 

\begin{theorem}
	\label{THM:Mismatch3}
	Under the conditional guessing problem under non-extensivity, let $G(X|Y)$ denote any arbitrary guessing strategy
	and let $\rho>0$, $q\in \mathbb{R}$. 
	Let $Q^{(G)}$ be the pmf associated with $G$ as obtained from \ref{THM:Mismatch2}. 
	Then
	\begin{eqnarray}
	 \left|R_q(P,G) - q\mathcal{RE}_{(\frac{q}{1+\rho},q)}(P,Q^{(G)}) \right|&\leq& \ln(1+\ln|\mathcal{X}|). 
	\label{EQ:mismatch_bound_redundancy}
	\end{eqnarray}
\end{theorem}
\noindent\textbf{Proof:}\\
Substituting $Q=Q^{(G)}$ in Theorem \ref{THM:Mismatch1}, 
and using (\ref{EQ:moment_inqE_Cond}) along with definition of $\mathcal{RE}_{(\alpha, \beta)}(P,Q)$,
one can easily deduce
$$
R_q(P, G) \leq q \mathcal{RE}_{(\frac{q}{1+\rho},q)}(P,Q^{(G)}) + \ln(1+\ln|\mathcal{X}|).
$$
On the other hand, from Theorem \ref{THM:Mismatch2} and (\ref{EQ:moment_inqE_Cond}), one can conclude
$$
R_q(P, G) \geq q \mathcal{RE}_{(\frac{q}{1+\rho},q)}(P,Q^{(G)}) - \ln(1+\ln|\mathcal{X}|),
\vspace{-.1in}$$
which completes the proof.
\hfill{$\square$}

Now, a optimal guessing strategy under source mismatch should work well for all possible true distributions $P\in\mathcal{P}$
and hence we should aim to minimize the worst redundancy measure given by $\sup\limits_{P\in\mathcal{P}}R_q(P,G)$.
However, from Theorem \ref{THM:Mismatch3}, it is expected that this optimal guessing strategy 
can be obtained from a pmf $Q$ that minimizes $\sup\limits_{P\in\mathcal{P}} q \mathcal{RE}_{(\frac{q}{1+\rho},q)}(P,Q)$,
or equivalently $\sup\limits_{P\in\mathcal{P}} \mathcal{RE}_{(\frac{q}{1+\rho},q)}(P,Q)$ if $q>0$.
We will now rigorously prove that it is indeed the case up to a factor of $\ln(1+\ln|\mathcal{X}|)$.
We start with the definition
\begin{eqnarray}
C_{q,\rho} = \min_Q \sup_{P\in\mathcal{P}} q\cdot \mathcal{RE}_{(\frac{q}{1+\rho},q)}(P,Q), ~~~~q\in\mathbb{R}, ~\rho > 0,
\end{eqnarray}
where we have assumed that the minimum exists and is attained, say at a pmf $Q^\ast$.
Then, we finally get an idea about how to find optimal guessing strategy and a bound of the worst-case redundancy value
in terms of $C_{q,\rho} $ which is presented in our final theorem below.
  
\begin{theorem}
\label{THM:Mismatch4}
Under the non-extensive conditional guessing problem, let $\rho>0$ and $q\in \mathbb{R}$ be such that
$C_{q,\rho}$ exists and is attained at $Q^\ast$. 
Then, for any arbitrary guessing strategy $G(X|Y)$, we have 
\begin{eqnarray}
\sup\limits_{P\in\mathcal{P}}R_q(P,G) \geq  C_{q, \rho} -  \ln(1+\ln|\mathcal{X}|). 
\label{EQ:mismatch_opt1}
\end{eqnarray}
Conversely, there exists a guessing strategy $\widetilde{G^\ast}(X|Y)$ that satisfies  
\begin{eqnarray}
\sup\limits_{P\in\mathcal{P}}R_q(P,\widetilde{G^\ast}) \leq  C_{q, \rho} +  \ln(1+\ln|\mathcal{X}|). 
\label{EQ:mismatch_opt2}
\end{eqnarray}
\end{theorem}
\noindent\textbf{Proof:}\\
For any arbitrary guessing strategy $G(X|Y)$, Theorem \ref{THM:Mismatch3} gives
$$
R_q(P, G) \geq q \mathcal{RE}_{(\frac{q}{1+\rho},q)}(P,Q^{(G)}) - \ln(1+\ln|\mathcal{X}|),
$$
Taking supremum over $P\in\mathcal{P}$, we get 
$$
\sup\limits_{P\in\mathcal{P}}R_q(P, G) \geq \sup\limits_{P\in\mathcal{P}}q \mathcal{RE}_{(\frac{q}{1+\rho},q)}(P,Q^{(G)}) 
- \ln(1+\ln|\mathcal{X}|)
\geq C_{q, \rho} -  \ln(1+\ln|\mathcal{X}|). 
$$

For the converse, note that, $C_{q, \rho} = \sup_{P\in\mathcal{P}}q \mathcal{RE}_{(\frac{q}{1+\rho},q)}(P,Q^\ast)$ by definition.
Take $\widetilde{G^\ast}= G_{Q^\ast}^\ast$. Then, as in the proof of Theorem \ref{THM:Mismatch3},
we get from Theorem \ref{THM:Mismatch1} and (\ref{EQ:moment_inqE_Cond}) that
$$
R_q(P, \widetilde{G^\ast}) \leq q \mathcal{RE}_{(\frac{q}{1+\rho},q)}(P,Q^\ast) + \ln(1+\ln|\mathcal{X}|).
$$
Taking supremum over $P\in\mathcal{P}$, we get 
$$
\sup\limits_{P\in\mathcal{P}}R_q(P, \widetilde{G^\ast}) 
\leq \sup\limits_{P\in\mathcal{P}} q \mathcal{RE}_{(\frac{q}{1+\rho},q)}(P,Q^\ast)  
+ \ln(1+\ln|\mathcal{X}|)
= C_{q, \rho} +  \ln(1+\ln|\mathcal{X}|). 
$$
This completes the proof.
\hfill{$\square$}

\section{Conclusion}

We have studied the guessing problem under non-extensive framework with $q$-normalized expectation. 
Our result generalizes the classical guessing results with usual expectation that formed the basis of Shannon coding theory.
Hence, it would be a natural follow-up work to apply our results to extend the Shannon coding theorem and related theory 
which will be helpful in order to develop and analyze more complex communication channel and related information theoretic problems. 
Our work opens up a new direction towards non-extensive information theory which we hope to study in more detail in our future work.

%
%
%


\end{document}